\documentclass[11pt,a4paper]{JHEP3}
\usepackage{amsfonts,ulem,bbold}   
\usepackage{amssymb,amsmath}

\newcommand{\be}{\begin{equation}}  
\newcommand{\ee}{\end{equation}}  
\newcommand{\bea}{\begin{eqnarray}}  
\newcommand{\eea}{\end{eqnarray}}  
\newcommand{\p}{\partial}  
\newcommand{\nn}{\nonumber}  
\newcommand{\wt}{\widetilde}  
\newcommand{\mA}{\mathbb A}
  
\newcommand{\mB}{\mathbb B}  
\newcommand{\mE}{\mathbb E}

\newcommand{\mX}{\mathbb X}

\newcommand{\one}{\mathbb 1} 
\newcommand{\I}{{\bf I}}

\newcommand{\tf}{{}^{{}^{{}_{3}}}\!\!f}
\DeclareMathOperator{\tr}{tr}  
\def\half{\frac{1}{2}}

\title{Ghost-free Massive Gravity with a General Reference Metric}  
   
\author{S. F. Hassan\\ Department of Physics \& The Oskar  
Klein Centre,\\ Stockholm University, AlbaNova University Centre,  
SE-106 91 Stockholm, Sweden \\ E-mail: \email{fawad@fysik.su.se}}  
  
\author{Rachel A. Rosen\\ Physics Department and Institute for
  Strings, Cosmology, and Astroparticle Physics,\\ 
Columbia University, New York, NY 10027, USA \\ E-mail:
\email{rar2172@columbia.edu}}   
  
\author{Angnis Schmidt-May\\ Department of Physics \& The Oskar Klein  
  Centre,\\ Stockholm University, AlbaNova University Centre, SE-106  
  91 Stockholm, Sweden \\ E-mail: \email
{angnis.schmidt-may@fysik.su.se}}  

\abstract{Theories of massive gravity inevitably include an auxiliary
  reference metric. Generically, they also contain an inconsistency
  known as the Boulware-Deser ghost. Recently, a family of non-linear
  massive gravity actions, formulated with a flat reference metric,
  were proposed and shown to be ghost free at the complete non-linear
  level. In this paper we consider these non-linear massive gravity
  actions but now formulated with a general reference metric. We
  extend the proof of the absence of the Boulware-Deser ghost to this
  case. The analysis is carried out in the ADM formalism at the
  complete non-linear level. We show that in these models there always
  exists a Hamiltonian constraint which, with an associated secondary
  constraint, eliminates the ghost. This result considerably extends
  the range of known consistent non-linear massive gravity theories.
  In addition, these theories can also be used to describe a massive
  spin-2 field in an arbitrary, fixed gravitational background. We
  also discuss the positivity of the Hamiltonian.}

\keywords{massive gravity}
   
\preprint{}   
 
\begin{document}    
\section{Introduction and summary}   
 
Generically, theories of massive gravity are plagued by the
Boulware-Deser (BD) ghost instability at the non-linear level
\cite{BD,BD2}.  Recently, significant progress has been made towards
constructing massive gravity theories that avoid this instability.  In
addition to the metric $g_{\mu\nu}$, theories of massive gravity
inevitably include another rank-2 symmetric tensor $f_{\mu\nu}$,
henceforth called the {\it reference metric}.  This is due to the fact
that the interaction terms that can be formed from the metric alone,
$\tr g=4$ and $\det g$, cannot be used to construct a mass term.  Most
of the recent work has focused on the case of a flat reference metric,
essentially, $f_{\mu\nu}= \eta_{\mu\nu}$.  In particular, a
two-parameter family of actions was proposed in \cite{dRG,dRGT} for
this case by demanding the absence of the BD ghost in what's known as
the decoupling limit.  One of these actions was demonstrated to be
ghost-free more generally at fourth order in perturbation theory in
\cite{dRGT}.  The full two-parameter family of actions were then shown
to be free of the BD ghost instability at the complete non-linear
level in \cite{HR3} based on the reformulation given in
\cite{HR1}. For complementary work see \cite{dRGT2,dRGT3,dRGT4}.

In this paper we consider non-linear massive gravity actions
constructed with a general $f_{\mu\nu}$ and extend the proof of the
absence of the BD ghost given in \cite{HR3} to this case. This
generalization is motivated by several considerations.  First, there
is no reason to insist that a theory of massive gravity always refer
to a flat reference metric. For example, one may also consider dS or
AdS metrics.  Second, forcing $f_{\mu\nu}$ to be flat constrains the
classical solutions of the metric $g_{\mu\nu}$.  For example,
non-Minkowski homogeneous and isotropic spacetimes were argued to be
excluded in \cite{D'Amico}.  A possible resolution of this problem is
to allow for more general $f_{\mu\nu}$ \cite{HR1}.  A third motivation
is that, from a theoretical standpoint, it is more satisfying to
promote $f_{\mu\nu}$ to a dynamical field with its own kinetic term
than to have a ``frozen-in" reference metric. The resulting theory
would resemble the bi-metric construction of \cite{ISS,SS}.  For a
dynamical $f_{\mu\nu}$ to be consistent, it is important to first
verify that the mass term which was ghost free for flat $f_{\mu\nu}$
remains so for a general $f_{\mu\nu}$.

It should be emphasized that although the discussion in this paper is 
formulated in the context of massive gravity, the analysis applies
equally well to generic massive spin-2 fields. For example,
$g_{\mu\nu}$ could also represent a neutral massive spin-2 meson in a
fixed gravitational background $f_{\mu\nu}$. 

At the linear level, massive gravity theories with a few simple
non-flat $f_{\mu\nu}$'s have already been considered. The linear
Fierz-Pauli theory of massive gravity \cite{FP1,FP2} in a flat
background has been generalized to linear massive gravities in de
Sitter and anti de Sitter spacetimes \cite{Higuchi,DW,Porrati,KMP},
and to FRW backgrounds \cite{Grisa,BDH1,BDH2}.  In these constructions
$f_{\mu\nu}$ plays the role of the background FRW metric.  Our work
explicitly shows that it is possible to construct non-linear
extensions of these theories that are free from the BD ghost
instability even for a general $f_{\mu\nu}$. Namely, the non-linear
ghost-free massive gravity actions proposed in \cite{dRGT} remain
ghost-free when constructed with respect to a general $f_{\mu\nu}$. 

In this work we consider the massive actions of \cite{dRGT} as
reformulated and extended to general $f_{\mu\nu}$ in \cite{HR1}. In
this reformulation, the two-parameter family is regarded as an
extension of a simpler, ``minimal'' massive action. Each free
parameter is associated with a higher level of non-linear complexity.
The simplicity of the minimal model is instrumental in constructing
the proof of the absence of the BD instability.  Moreover, for this
model the constraint equations can be solved explicitly, making it
possible to study issues such as the positivity of the Hamiltonian.
Once the proof of the absence of the BD ghost is constructed for the
minimal model, we find that the exact same construction holds for the
more complicated two-parameter family of actions.  

Our analysis is based on the ADM formulation of gravity \cite{ADM}.
In the ADM language, the BD ghost is a consequence of the absence of
the Hamiltonian constraint.  We show that in the models considered
here, such a constraint exists.   With an associated
secondary constraint (see \cite{HR6}), this is enough to eliminate the BD
ghost mode.

The paper is organized as follows: Section 2 starts with a review of
non-linear massive gravity with a general reference metric
$f_{\mu\nu}$.  We discuss the Boulware-Deser ghost problem in
non-linear massive gravity and present precise criteria for avoiding
it.  We then review the specific two-parameter family of actions
considered in this paper.  In section 3 we show the absence of the BD
ghost in the minimal massive action by obtaining the Hamiltonian
constraint and arguing for the existence of an associated secondary
constraint. We then discuss the positivity of the Hamiltonian.  In
section 4, the proof of the absence of the BD ghost is extended to the
complete two-parameter family of massive actions and the Hamiltonian
constraint is determined.  The results are briefly discussed in
section 5.  In the appendix we review the ghost issue in linear and
non-linear massive gravities, including the original analysis of
Boulware and Deser \cite{BD}.

\section{Review of massive gravity and the Boulware-Deser ghost problem}

In this section we discuss the ghost problem in massive gravity,
reviewing the Boulware-Deser argument and the caveat by which it can
be avoided.  We also review the two-parameter family of potentially
ghost-free massive gravity actions formulated with respect to a
general $f_{\mu\nu}$.
  
\subsection{General structure of non-linear massive gravity}

A generic covariant massive gravity action for the metric $g_{\mu\nu}$
is obtained by adding a non-derivative potential term $V(g^{-1}f)$ to
the Einstein-Hilbert action {\cite{BD}},
\be
S_m=M_p^2\int d^4 x\sqrt{-g}\left[R(g)- m^2 \,V(g^{-1}f) \right] \, .
\label{Sm-generic}
\ee
$f_{\mu\nu}$ is a non-dynamical rank 2 tensor that is needed to
construct generally covariant, non-derivative functions of the
metric. The coupling of the metric $g_{\mu\nu}$ to matter is taken to
be the same as in GR in order to preserve the weak equivalence
principle. Below, we will have more to say about the role of
$f_{\mu\nu}$.

Such a generic non-linear massive gravity action (\ref{Sm-generic})
typically contains a ghost, i.e., a physical mode with negative
kinetic energy which in the quantum theory results in negative
probability states.  The origin of the ghost is easy to understand
(see the appendix for a detailed review of the ghost problem).  In
general relativity, a scalar component of the metric is potentially a
ghost, but is eliminated by the equations of motion. The addition of a
potential energy term to the GR action generally results in this
component becoming an independent dynamical degree of freedom. Then
the theory has six propagating modes: the five polarizations of the
massive spin-2 graviton and the ghost.

At the linear level, the ghost problem is avoided by the mass term
proposed by Fierz and Pauli \cite{FP1,FP2}.  A necessary (but not
sufficient) requirement for the action (\ref{Sm-generic}) to be
ghost-free is that, when expanded to quadratic order in metric
fluctuations $h_{\mu\nu}=g_{\mu\nu}-\bar g_{\mu\nu}$, the potential
$V$ should reproduce the Fierz-Pauli (FP) mass term in the background
$\bar g_{\mu\nu}$, provided one takes $\bar g_{\mu\nu}=f_{\mu\nu}$,
\be
S_{FP}=-\frac{1}{4}M_p^2 m^2 \int d^4x \sqrt{-\bar g}\left[ 
h^\mu_\nu\, h_\mu^\nu\,- (h^\mu_\mu)^2\,\right]\,,
\label{FP-bg}
\ee
where, $h^\mu_\nu=\bar g^{\mu\rho}h_{\rho\nu}$.  At the linear level,
this particular choice of relative coefficients between the terms
decouples the ghost by making it infinitely massive.  However, it was
shown by Boulware and Deser \cite{BD,BD2} that the ghost sixth mode
generally reappears at the non-linear level, leading to speculation
that a non-linear theory of massive gravity may not exist.  We will
discuss this further below.

For the purpose of identifying the massive excitations, $f_{\mu\nu}$
must be equated to a background metric, $f_{\mu\nu} =\bar
g_{\mu\nu}$. Then the background field equations reduce to the GR
equations with a shifted cosmological constant and $f_{\mu\nu}$ is a
solution for a given source, say, $\bar T_{\mu\nu}$. For this reason
$f_{\mu\nu}$ is often referred to as a ``background metric''. But,
given $f$, the non-linear theory will also have classical solutions in
which $g$ differs appreciably from $f$ in some regions of spacetime,
see for example, \cite{BDZ1,TMN,Koyama1,Koyama2}. Any such solution
can be regarded as a background $\bar g$ with fluctuations
$h_{\mu\nu}'$ around it, although the action for these fluctuations
may no longer have the Fierz-Pauli form (\ref{FP-bg}).  Thus at the
non-linear level, one could consider fluctuations around background
metrics other than $f_{\mu\nu}$.  For this reason we refer to
$f_{\mu\nu}$ as the ``reference metric'', rather than a background
metric.  The physical metric of spacetime is still $g_{\mu\nu}$.

Of the ten components of $f_{\mu\nu}$, four are gauge degrees of
freedom, removable by gauge fixing general coordinate
transformations. This is made explicit in the parameterization, 
\be
f_{\mu\nu}=\frac{\p\phi^a}{\p x^\mu}\bar f_{ab}\frac{\p\phi^b}{\p x^\nu}\, .
\label{f-phi}
\ee
The $\phi^a$ are interpreted as St\"uckelberg fields or as Goldstone
modes associated with the breaking of general covariance
\cite{AGS}. The remaining six components contained in $\bar f$ are
non-dynamical.  Possible choices for $f_{\mu\nu}$ are:

{\it Flat reference metric:} Most of the recent work on massive
gravity has focused on $\bar f_{\mu\nu}=\eta_{\mu\nu}$.  In the
unitary gauge this gives $f_{\mu\nu}=\eta_{\mu\nu}$.  For this choice,
(\ref{FP-bg}) is the original ghost free Fierz-Pauli mass term
\cite{FP1,FP2} for metric fluctuations around flat spacetime. Later,
the generic instability of the non-linear theory (\ref{Sm-generic})
was shown by Boulware and Deser \cite{BD,BD2} for this case, although
their analysis also applies to general $f$. The actions recently
proposed in \cite{dRG,dRGT} also belong to this class, where the
fields $\phi^a$ (\ref{f-phi}) played an important role in the
construction.  The absence of the Boulware-Deser instability at the
complete non-linear level was proved for these actions in \cite{HR3}.

{\it FRW reference metric:} The quadratic action (\ref{FP-bg}) is also
known to be free of the Boulware-Deser ghost instability when
$f_{\mu\nu}$ is a de Sitter or anti de Sitter metric
\cite{Higuchi,DW,Porrati,KMP}, or more generally, an FRW
\cite{Grisa,BDH1,BDH2} metric\footnote{It turns out that, in some
  regions of parameter space, these theories may suffer from
  instabilities quite distinct from the Boulware-Deser problem, even
  at the linear level.  However, these do not necessarily reflect an
  inconsistency of the theory \cite{Grisa,BDH1,BDH2}.}.  However,
consistent non-linear extensions of such quadratic actions had so far
remained undetermined. 

{\it General non-dynamical reference metric:} In this paper we
demonstrate the consistency of the non-linear massive actions proposed
in \cite{dRGT}, when extended to general $f_{\mu\nu}$ \cite{HR1}.
Such an extension is not only natural, but is also necessary to obtain
a larger and potentially more viable class of solutions.

{\it Dynamical reference metric:} It is appealing to complete the
theory by including dynamics for $f_{\mu\nu}$ \cite{ISS,SS}. That this
can be done consistently in the context of bi-metric theories of
gravity will be demonstrated in an accompanying work \cite{HR5}.

\subsection{The ADM formulation of general relativity}

The physical content of gravity and its propagating modes are 
easily identified in the ADM formulation \cite{ADM} which is based on
a $3+1$ decomposition of the metric,   
\be
N \equiv (-g^{00})^{-1/2}\,, \qquad N_i \equiv g_{0i}\,, \qquad
\gamma_{ij}\equiv g_{ij}\, .  \\ 
\ee
The $N$ and $N_i$ are the lapse and shift functions respectively. In
this parameterization,
\be
g^{\mu\nu}=N^{-2}\left(\begin{array}{cc}\quad -1 \quad & N^j\\[.2cm] 
 N^i & N^2\gamma^{ij}- N^iN^j \end{array}\right) \,,
\label{gADM}
\ee
where, $N^j=\gamma^{jk}N_k$ and $\gamma^{ij}\gamma_{jk}=\delta^i_k$.
Denoting the momentum canonically conjugate to $\gamma_{ij}$ by
$\pi^{ij}$, the Einstein-Hilbert action in terms of these variables
becomes (we ignore all boundary terms in what follows),
\be
S=M_p^2\int d^4x \left[\pi^{ij}\p_t\gamma_{ij}+ N R^0+ N_i R^i\,\right] \, .
\label{S-GR-ADM}
\ee
The $R^{\mu}$ are functions of $\gamma_{ij}$ and $\pi^{ij}$
but are independent of the $N_\mu=(N,N_i)$, 
\be
R^0=\sqrt{\det \gamma}\left[R(\gamma)+\frac{1}{\det\gamma}(\half\pi^2-
  \pi^{ij}\pi_{ij}) \right]\,,\qquad 
R^i=2\sqrt{\det\gamma}\,\nabla_j \left(\frac{\pi^{ij}}
{\sqrt{\det\gamma}} \right)\,.\qquad  
\ee
The six components of $\gamma_{ij}$ are potentially propagating modes
in the sense that their equations of motion obtained from
(\ref{S-GR-ADM}), as well as those for their conjugate momenta
$\pi^{ij}$, involve time derivatives (so that the Euler-Lagrange
equations for $\gamma_{ij}$ are second order in time).  Since a single
propagating mode involves a field component and its canonically
conjugate momentum, the six potentially propagating modes are
described by the 12 functions $(\gamma_{ij},\pi^{ij})$.  However, in
the theory defined by (\ref{S-GR-ADM}) not all of these are
independent.  To see this, note that the $N_\mu$ appear linearly as
Lagrange multipliers, hence their equations of motion are four
constraints (the ``Hamiltonian'' and ``momentum'' constraints) on the
remaining fields,
\be
R_0(\gamma, \pi) =0\,,\qquad  R_i(\gamma, \pi) =0 \, .
\ee
These constraints can be used along with the four general coordinate
transformations to eliminate eight of the 12 functions, in favor of
two remaining pairs. These pairs are the two propagating modes of GR,
describing the two polarization states of the massless graviton at the
non-linear level.  In particular, the scalar ghost is not part of the
physical spectrum \cite{ADM}.

Of the 12 equations of motion for $(\gamma_{ij},\pi^{ij})$, four
reduce to Bianchi identities while another four determine the
$N_\mu$. The remaining four equations describe the propagating modes.

\subsection{The Boulware-Deser ghost}

Boulware and Deser used the ADM formalism to study the physical
content of non-linear massive gravity and argued that, generically,
massive gravity has six propagating modes. The sixth mode is the ghost
that was avoided in the linear FP action but reappears at the
non-linear level \cite{BD,BD2}.  They also found that the non-linear
theory had a pathological non-positive Hamiltonian. Let us summarize
their analysis here.

In ADM parameterization, the massive gravity action (\ref{Sm-generic})
becomes,  
\be
S_m=M_p^2\int d^4 x\left[ \pi^{ij}\p_t \gamma_{ij}+N R^0+N_i R^i-
m^2 V'(\gamma,N,N_i, \bar f_{\mu\nu}) \right]\,,
\label{Sm-genADM}
\ee
where $V'=\sqrt{\det\gamma}\,NV$ and coordinate transformations have
been used to set $f_{\mu\nu}=\bar f_{\mu\nu}$ (\ref{f-phi}). Since
$V'$ is a non-linear function of the $N_\mu$, the lapse and shift are
no longer Lagrange multipliers and their equations of motion,
\be
R^\mu(\gamma,\pi)=m^2 \, V^\mu(\gamma,N,N_i,\bar f)\,,\qquad
\mbox{\rm with}\qquad V^\mu \equiv \frac{\p V'}{\p N_\mu}\,,
\label{Rmu-gen}
\ee
no longer constrain $\gamma_{ij}$ and $\pi^{ij}$.  Instead, these
equations can be solved for the $N_\mu$ in terms of
($\gamma_{ij},\pi^{ij}$).  After eliminating the $N_\mu$ in this way,
one is left with twelve equations for the twelve dynamical variables
($\gamma_{ij},\pi^{ij}$), hence the theory contains six propagating
modes.  In particular the sixth, ghost mode that was avoided in the
linear FP theory has re-emerged as a propagating mode.  Boulware and
Deser \cite{BD} argued that, since in massive gravity $V'$ is always
non-linear in the $N_\mu$, the sixth mode cannot be avoided.

As an explicit example of a non-linear mass term, 
\cite{BD} considered the FP mass (\ref{FP-bg}) with
$f_{\mu\nu}=\eta_{\mu\nu}$, where now $h_{\mu\nu} =
g_{\mu\nu}-\eta_{\mu\nu}$ is no longer treated as a small
fluctuation. This analysis is reviewed in the appendix. They found that
in the linear approximation the Hamiltonian constraint eliminates the
ghost, hence the linear FP theory is indeed consistent.  However, at
the non-linear level, for the mass term considered, there is no such
constraint.  Thus, to reiterate, the Boulware-Deser ghost instability
is due to the loss of the Hamiltonian constraint at the non-linear
level. Finally, in \cite{BD,BD2} it was concluded that,
\begin{itemize}
\item The massive theory has six rather than five degrees of freedom,
  and hence contains a ghost.  
\item The Hamiltonian of the massive theory is not positive definite.   
\item In the limit $m\rightarrow 0$ the Hamiltonian diverges, hence
  this limit does not exist. 
\end{itemize}

There is, however, a caveat in the arguments of Boulware and Deser.
As pointed out in \cite{dRGT}, avoiding the Boulware-Deser instability
does not, in fact, strictly require linearity of the theory in the
$N_\mu$.  Rather, it is enough that one combination of the four
$N_\mu$ equations of motion (\ref{Rmu-gen}) becomes a constraint on
the $(\gamma_{ij},\pi^{ij})$. Based on this observation, we formulate
the criteria for the absence of the ghost in the following
subsection.

\subsection{Criteria for the absence of the Boulware-Deser ghost}
\label{criteria}

The caveat in the Boulware-Deser argument can be stated as
follows. While the potential $V(g^{-1}f)\equiv V(N,N_i,\gamma,f)$ is
 a non-linear function of the $N_\mu$, suppose there exist
potentials $V$ for which the $N_\mu$ equations of motion depend only
on three combinations of the $N_\mu$,
\be
n_r=n_r(N, N^i,\gamma)\,, \qquad r=1,2,3 \, .
\label{n_r}
\ee
That is, the $N_\mu$ equations (\ref{Rmu-gen}) take
the generic form,  
\be
R^\mu(\gamma,\pi)=m^2 \, \wt V^\mu(\gamma,n_r)\,.
\label{Rmu-cav}
\ee
Then, in principle, three of these equations can be used to determine
the $n_r$ in terms of ($\gamma,\pi$). Substituting the result into the
remaining equation gives a constraint on the ($\gamma,\pi$) that may
have the right form to eliminate the ghost field. Finally, one also
needs a second constraint to eliminate the variable canonically
conjugate to the ghost field.

The linear Fierz-Pauli theory
 is linear in the lapse but not in the shift.  Thus the $N$
equation of motion provides a modified Hamiltonian constraint while
the $N_i$ equations are not constraints, but determine the $N_i$ in
terms of the $(\gamma,\pi)$. It is natural to expect that this feature
extends to a ghost-free non-linear theory, especially since the ghost
is a scalar.  Thus the functions $n_i(N,N_j,\gamma)$ can be regarded
as the counterparts of $N_i$ in the massive theory, consistent with
the 3-dimensional general covariance maintained in the ADM
formulation. Assuming that the functional relationship is invertible
(as it should be), one can determine the $N_i$ as functions of the
$n_i$:  $N_i(N,n_j,\gamma)$. Then the massive action (\ref{Sm-genADM})
can be expressed in terms of the combinations $n_i$,
\be 
S[N, N_i]=\wt S[N, n_j(N,N_i,\gamma)]\,. 
\ee
Now consider the $N_\mu$ equations of motion,   
\be
\frac{\delta S}{\delta N_i}\equiv\frac{\delta\wt S}{\delta n_j} 
\Big\vert_N\, 
\frac{\delta n_j}{\delta N_i}=0\,, \qquad
\frac{\delta S}{\delta N}\equiv\frac{\delta\wt S}{\delta N}\Big\vert_{n}+  
\frac{\delta\wt S}{\delta n_j}\Big\vert_N\,\frac{\delta n_j}{\delta N} 
=0 \, ,
\label{eomNNi}
\ee
where the subscript on the vertical bar indicates the variable held
fixed in the process of variation. This leads to the equivalent equations, 
\be
\frac{\delta\wt S}{\delta n_j}\Big\vert_N = 0\,, \qquad
\frac{\delta\wt S}{\delta N}\Big\vert_n =0\,,
\label{eomNni}
\ee 
which are linear combinations of the $N_\mu$ equations. Based on these
equations one can now formulate a nested set of criteria for the
existence of a Hamiltonian constraint.
\begin{enumerate}
\item As described above, for a constraint to exist, the $n_i$
  equations of motion should depend on $N_\mu$ only through the three
  combinations $n_i$. Then they can be used to determine the $n_i$ in
  terms of $(\gamma_{ij},\pi^{ij})$. 
\item The $N$ equation also must involve only the $n_i$ and be
  independent of $N$ so that, given the $n^i$ solution, it becomes a
  constraint on $(\gamma_{ij},\pi^{ij})$.  For this to be the case,
  the massive action in the form $\wt S$, i.e., when regarded as a
  functional of $N$ and $n_i$, must be linear in $N$.
\item The action $\wt S$ also contains the term $N_i R^i$ where $N_i =
  N_i(N,n_j,\gamma)$.  Linearity of $\wt S$ in $N$ then implies that
  the expression for $N_i$ in terms of $n_j$ must be linear in $N$.
\end{enumerate}
Note that the minimal coupling of the metric to matter is linear in
the lapse and shift.  If this were not the case, the constraints of
even massless GR would be violated.  Thus criteria specified here are
not modified by the presence of the minimal matter coupling.

In this paper we show that, for the massive gravity theories described
in the next subsection, once requirements 2 and 3 are satisfied, then
1 follows automatically. This guarantees the existence of a
Hamiltonian constraint associated with the $N$ equation of motion. We
also argue for the existence of a non-trivial secondary constraint
(subsequently proven in \cite{HR6}) as, simply, the non-linear
extension of the known secondary constraint in the linear FP
theory (see, e.g., \cite{Hinterbichler}). These two constraints
eliminate the canonical pair corresponding to the Boulware-Deser
ghost, reducing the number of propagating modes from six to five.

\subsection{Non-linear massive gravity actions with general $f_{\mu\nu}$} 

In principle, the above criteria might be used to construct a theory
of massive gravity that does not suffer from the Boulware-Deser
instability. In practice, this was not so straightforward.\footnote{In
  hindsight, these criteria are powerful enough that they can
  determine the complete form of the non-linear action as will be
  discussed elsewhere \cite{HRS2}.} Potentially ghost free actions
were first constructed for $\bar f_{\mu\nu}=\eta_{\mu\nu}$ following a
very different perturbative argument.  It was observed in \cite{AGS}
that the $\phi^a$ in (\ref{f-phi}) are the Goldstone bosons associated
with the breaking of general covariance by the mass term.  Then for
$\bar f_{\mu\nu}=\eta_{\mu\nu}$ and $g_{\mu\nu} =
\eta_{\mu\nu}+h_{\mu\nu}$, an analogy with the Goldstone-vector boson
equivalence theorem in gauge theory implies that, in the high energy
limit, the dynamics of massive gravity is mirrored in the dynamics of
the Goldstone sector, particularly, in the ``longitudinal'' mode of
the $\phi^a$ fluctuations. Thus the ghost of massive gravity appears
as a ghost in this longitudinal mode. Being a scalar field in flat
spacetime, this is a much easier setup to investigate. One may attempt
to constrain $V$ using this correspondence \cite{CNPT}.

The breakthrough came with the work of de Rham and Gabadadze
\cite{dRG} and de Rham, Gabadadze and Tolley \cite{dRGT} who used this
approach to construct a two-parameter family of massive actions for
$\bar f_{\mu\nu}=\eta_{\mu\nu}$.  The two free parameters are denoted
by $\alpha_3$ and $\alpha_4$.  These actions were shown to be ghost
free in this high energy limit, the ``decoupling limit''.  To go
beyond the decoupling limit, \cite{dRGT} carried out an ADM analysis
of the $\alpha_3=\alpha_4=0$ model to quartic order in the metric
perturbations and demonstrated the absence of the Boulware-Deser ghost
to that order.

The presentation of the actions given in \cite{dRGT} is convenient for
a perturbative analysis.  However, the expressions that multiply the
parameters $\alpha_n$ contain mixtures of terms with different levels
of non-linear complexity. To demonstrate the absence of the
Boulware-Deser ghost at the full non-linear level it is helpful to use
the reformulation of \cite{HR1} in which different levels of
non-linearity are disentangled. Using this reformulation, the absence
of the Boulware-Deser ghost at the non-linear level was proven for
$\bar f_{\mu\nu}=\eta_{\mu\nu}$ in \cite{HR3}.

In this paper we extend the ghost analysis to any general,
non-dynamical $f_{\mu\nu}$ using the presentation of massive gravity
actions given in \cite{HR1}.  Let us briefly review this formulation.
The basic building block of non-linear massive gravity is the
square-root matrix $\sqrt{g^{-1}f}$ \cite{dRGT,HR1}, where
$\sqrt{g^{-1}f} \sqrt{g^{-1}f} = g^{\mu \lambda}f_{\lambda \nu}$.  The
terms appearing in the massive action are identified as elementary
symmetric polynomials of the eigenvalues of this matrix.  They sum up
to a ``deformed determinant'' (for related ideas see
\cite{TMN,Koyama2}). The antisymmetry property of this structure
allows one to generalize the reference metric from flat to any
$f_{\mu\nu}$ and still remain within the same 2-parameter family of
actions.
  
The simplest non-linear massive action with zero cosmological constant
is given by \cite{HR1},  
\be
\label{actmin}  
S_{min}=M_p^2\int d^4x\sqrt{-g}\left[\,R 
-2  m^2\,\left(\tr \sqrt{g^{-1}f}-3\right)\, \right]\, .
\ee  
We will refer to this as the minimal massive action. The most general
non-linear massive action can be written as  
\be
\label{act2}  
S=M_p^2\int d^4x\sqrt{-g}\,\bigg[R +2m^2 \sum_{n=0}^{3} \beta_n\,
  e_n(\sqrt{g^{-1} f})\bigg]\,.
\ee 
The $e_k(\mX)$ are elementary symmetric polynomials of the eigenvalues
of $\mX$.  For a generic $4\times 4$ matrix they are given by, 
 \bea
\label{ek}   
e_0(\mX)&=& 1  \, , \nonumber \\  
e_1(\mX)&=& [\mX]  \, ,\nonumber \\  
e_2(\mX)&=& \tfrac{1}{2}([\mX]^2-[\mX^2]), 
\label{e_n}  \\  
e_3(\mX)&=& \tfrac{1}{6}([\mX]^3-3[\mX][\mX^2]+2[\mX^3])
\, ,\nonumber \\   
e_4(\mX)&=&\tfrac{1}{24}([\mX]^4-6[\mX]^2[\mX^2]+3[\mX^2]^2   
+8[\mX][\mX^3]-6[\mX^4])\, ,\nonumber \\  
e_k(\mX)&=& 0 \qquad {\rm for} \quad k>4 \, , \nonumber
\eea 
where the square brackets denote the trace.  Of the four $\beta_n$,
two combinations are related to the mass and the cosmological
constant, while the remaining two combinations are free
parameters. Setting the cosmological constant to zero and the
parameter $m$ as the mass, the four $\beta_n$ are parameterized in
terms of the $\alpha_3$ and $\alpha_4$ of \cite{dRGT} as (for $n = 0,
\ldots, 4$),
\be
\beta_n=(-1)^n\left(\half (4-n)(3-n)-(4-n)\alpha_3+\alpha_4\right)\,,
\ee 
The minimal action corresponds to $\beta_2=\beta_3=0$ supplemented by 
$\beta_0=3$, $\beta_1=-1$ to get a zero cosmological constant
contribution from the potential term. We will start with this minimal
theory in the analysis that follows.

\section{Absence of the BD ghost in the minimal massive action} 

In this section we show that the minimal non-linear massive gravity
action (\ref{actmin}) satisfies the criteria outlined in section
\ref{criteria} and thus there exists a Hamiltonian constraint on the
dynamical variables. In addition, we argue for the existence of an
associated secondary constraint. Thus this action does not suffer from
the Boulware-Deser ghost instability. We solve the constraints
explicitly and discuss the positivity of the Hamiltonian.

\subsection{Enforcing the criteria for the existence of the
  Hamiltonian constraint} 
In the ADM formulation the Lagrangian for the minimal massive action
(\ref{actmin}) becomes,
\be
{\cal L}_{min}= \pi^{ij}\partial_t \gamma_{ij} + N R^0 +N^i
R_i-2m^2\sqrt{\det\gamma}\, N\,\left(\tr \sqrt{g^{-1}f}-3\right) ,
\label{LminADM}
\ee
where, using the parameterization (\ref{gADM}) for $g^{\mu\nu}$, one
gets,
\be
\label{ADM}
N^2\,g^{-1}f = 
\left( \begin{array}{ccc}
-f_{00}  +N^l f_{l0}  &~~& -f_{0j}+N^l f_{lj} \\[.1cm]
N^2 \gamma^{il}f_{l0} -N^i(-f_{00}  +N^l f_{l0} )&& N^2
\gamma^{il}f_{lj} -N^i(-f_{0j}+N^l f_{lj})
\end{array} \right) \, .
\ee
Since the action contains the square root of this matrix, it is highly
non-linear in the $N_\mu$. Hence it could potentially propagate a
ghost sixth mode according to the Boulware-Deser argument. But, as
discussed in section \ref{criteria}, this can be avoided if the four
$N_\mu$ equations of motion happen to depend only on three
combinations of the lapse and shift, say $n^i(N_\mu)$, leaving a
single constraint to eliminate the sixth, ghost mode.

We show now that this is indeed the case for the minimal action
(\ref{LminADM}). First, we identify the appropriate functions $n^i$.
This is achieved by imposing the criteria for the absence of ghost
discussed in section \ref{criteria}. In fact, we will only have to
impose criteria 2 and 3. Then 1 follows automatically.

Criterion 3 requires that the expression for $N_i$ in terms of the
$n_i$ be linear in $N$,   
\be
N^i = c^i + N d^i\,.
\label{Nini}
\ee
The $c^i$ and $d^i$ are functions of $n^i$ and $\gamma_{ij}$ but are
independent of $N$. They will be determined in what follows by
demanding that the action, when written in terms of $n_i$ and $N$,
must be linear in $N$. Using (\ref{Nini}), (\ref{ADM}) takes the form 
\be
N^2\,g^{-1}f= \mE_0 + N \, \mE_1 + N^2\, \mE_2    
\label{ADMn}
\ee
where the matrices $\mE_0$, $\mE_1$ and $\mE_2$ are independent of
$N$.  To write them compactly, define, 
\be
a_\mu = -f_{0\mu} + c^l f_{l \mu} \,.
\label{amu}
\ee
Then one gets,
\be
\mE_0=\left( \begin{array}{ccc}
a_0  &~& a_j \\[.1cm]
-a_0 c^i &&  -c^ia_j 
\end{array} \right)
\,,\qquad 
\mE_2=\left( \begin{array}{ccc}
0 &~~& 0 \\[.1cm]
(\gamma^{il}-d^id^l) f_{l0} && (\gamma^{il}-d^id^l) f_{lj} 
\end{array} \right) \,, 
\label{E0E2}
\ee
and,
\be
\mE_1=\left( \begin{array}{ccc}
d^l f_{l0}&~~&  d^l f_{lj}\\[.1cm]
-(d^l f_{l0}c^i+a_0d^i)&&-(c^id^l f_{lj}+d^i a_j) 
\end{array} \right) \,.
\label{E1}
\ee

Criterion 2 of section \ref{criteria} requires that the mass term,
when written in terms of $N$ and $n_i$, must be linear in $N$. For the
minimal massive action, this is satisfied if the matrix $\sqrt{g^{-1}f}$
has the following form\footnote{Note that this is more restrictive
  than requiring the linearity of the $\tr(N\sqrt{g^{-1}f})$ in $N$,
  but leads to simple systematics that satisfy requirement 1
  automatically.},
\be
N \sqrt{g^{-1}f} = \mA +N \mB \, ,
\label{AB}
\ee
where the matrices $\mA$ and $\mB$ are independent of $N$. Demanding
that this expression (\ref{AB}) be consistent with $N^2g^{-1}f$ as
given by (\ref{ADMn}), determines $\mA$ and $\mB$ as well as $c^i$ and 
$d^i$. Explicitly, comparing (\ref{AB}) and (\ref{ADMn}) gives,
\be 
\mA^2=\mE_0\,, ~~~~\mB^2=\mE_2\, ,~~~~{\rm and}~~~~ \mA\mB+\mB\mA=\mE_1 \, .
\label{AB2}
\ee
Let us consider the first two equalities in (\ref{AB2}).  Using
(\ref{E0E2}) it is easy to verify that these imply, 
\be
\mA = \frac{1}{\sqrt{x}}\left( \begin{array}{ccc}
a_0  &~~& a_j \\[.1cm]
-a_0c^i && -c^ia_j
\end{array} \right) \,,
\qquad
\mB =\sqrt{x} \left( \begin{array}{ccc}
0&~~&0 \\[.1cm]
D^i_{~k} (\tf^{-1})^{kl} f_{l0} &&  D^i_{~j}
\end{array} \right) \, .
\label{mAmB}
\ee
Here $\tf_{ij}\equiv f_{ij}$ and we have introduced,
\be
x \equiv a_0-c^l a_l\,,\qquad 
\sqrt{x}\, D^i_{\,\,j} \equiv \sqrt{(\gamma^{il}-d^id^l) f_{lj}} \, .
\label{xD}
\ee
The expression for $\mA=\sqrt{\mE}_0$ follows since $\mE_0$ is a
projection operator, $\mE_0^2 =x \mE_0$, hence $\mA=\mE_0/\sqrt{x}$. 
In the expression for $\mB$, the square-root matrix $D$ is defined
with an extra factor of $\sqrt{x}$ for later convenience.

Before proceeding further, we note a very important property of the
matrix $D$. According to (\ref{xD}) it has the form
$D=\sqrt{S\,\,\tf}$ where both $S$ and $\tf$ are symmetric matrices.
By rewriting $D$ as $\sqrt{1+(S\,\,\tf-1)}$ and then expanding in
powers of $(S\,\,\tf-1)$, it becomes obvious that $(\tf D)^T = \tf D$.
In terms of components, this means
\be
\label{fD}
f_{ik}D^k_{~j} = f_{jk}D^k_{~i} \, .
\ee
This identity will be used often in the following analysis.

Now let us consider the third equality in (\ref{AB2}). Using
(\ref{mAmB}), one can compute $\mA\mB+\mB\mA$ and compare the result
with $\mE_1$ in (\ref{E1}). Using (\ref{amu}) and the property of $D$
given in (\ref{fD}), one obtains the following relation between $c^i$
and $d^i$,
\be
d^i = D^i_{~k} \left[ c^k-(\tf^{-1})^{kl} f_{l0}\right] \, .
\label{cd}
\ee
Guided by the case of flat $f_{\mu\nu}$ considered in \cite{HR3}, we
introduce the variables $n^i$ so that\footnote{This choice simplifies
  some of the equations, but is not unique. For instance, we could
  have also chosen $n^i=c^i$. For a different choice, see section
  \ref{hats} below.}, 
\be
n^k= c^k -(\tf^{-1})^{kl} f_{l0} \, .
\ee
Then, (\ref{cd}) reduces to,
\be
d^i = D^i_{\,k}\,n^k \, .
\label{d}
\ee
Substituting for $d^i$ in (\ref{xD}) gives a matrix equation for $D$, 
\be
\sqrt{x}\,D =\sqrt{(\gamma^{-1}-D n n^T D^T)\,\tf} \, ,
\label{D}
\ee 
which will be solved below in terms of $n^i$. Thus $c^i$, $d^i$ and
$D^i_{~j}$ can be determined entirely in terms of the $n^i$ and
$\gamma_{ij}$. This proves that indeed there exist modified shift
variables $n^i$ in terms of which $N\sqrt{g^{-1}f}$ is linear in $N$
and given by (\ref{AB}). 

In the proof that follows, we need only the condition (\ref{D}), and
not its explicit solution. However, it is important that this solution
exists and can be used to show that the relation (\ref{NnD}) is
invertible. Thus we take a moment to derive the solution. Squaring
both sides of (\ref{D}) and moving the $D$-dependent terms to one side
gives, after using (\ref{fD}),
\be
\label{Dsol1}
D^i_{~l}\, Q^l_{~j}\,D^j_{~k}=\gamma^{ij} f_{jk}\,,\quad
\text{with,}\quad 
Q^l_{~j} = x\,\delta^l_{~j}+n^l n^m f_{mj} \, .
\ee
On multiplying both sides by $Q$, this becomes $(D\,Q)^2 =
(\gamma^{-1}\,\tf)Q$. Taking the square root and rearranging gives,   
\be
D=(\sqrt{\gamma^{-1}\,\tf Q})\, Q^{-1}\,.
\label{SolnD}
\ee
The inverse matrix $Q^{-1}$ is easily obtained by noting that $(n
n^T\,\tf)^2 = (n^T\,\tf n)n n^T\,\tf$.  Then one finds 
\be
Q^{-1}=\frac{1}{x}(1- M^{-2}\, n n^T\tf)\,,
\label{Minv}
\ee
where $M^2=-f_{00}+ f_{0k} (\tf^{-1})^{kl} f_{l0}$ is the lapse
function of $f_{\mu\nu}$. Equations (\ref{SolnD}) and (\ref{Minv})
give the explicit solution for $D$ in terms of $n^i$.

Before moving on, let us point out that our final expressions
naturally involve the ADM parameters of $f_{\mu\nu}$. In analogy with
the ADM parameterization of $g_{\mu\nu}$, we define,
\be
M \equiv (-f^{00})^{-1/2}\,, \qquad M_i \equiv f_{0i}\,, \qquad
\tf_{ij}\equiv f_{ij}\, . \\ 
\ee
We also define $M^i=(\tf^{-1})^{ij}M_j$. Then, the variables $a_\mu$
defined in (\ref{amu}) become,
\be
a_0=M^2+n^lM_l\,,\qquad a_i=\tf_{il}n^l\,,
\ee
and in terms of the lapse $M$, the $x$ of (\ref{xD}) is simply, 
\be
x=M^2 - \,n^k\, f_{kl}\,n^l\,.
\label{x}
\ee 

To recapitulate, we have identified three variables $n^i$ such that
the Lagrangian (\ref{LminADM}) written in terms of these variables is
linear in $N$ and hence satisfies criteria 2 and 3 for the existence
of a Hamiltonian constraint, as outlined in section
\ref{criteria}. The functions $n^i$ are related to the lapse and shift
variables of $g_{\mu\nu}$ and $f_{\mu\nu}$ through\footnote{For
  $f_{\mu\nu}=\eta_{\mu\nu}$, in \cite{dRGT} a perturbative relation
  between $N_i$ and $n_i$ was used to demonstrate the absence of the
  BD ghost to quartic order in the fluctuations. To compare this
  perturbative relation with our result we expand (\ref{NnD}) to
  quartic order around flat spacetime using (\ref{SolnD}). After
  lowering the indices and rescaling $n_i$ by 2, one gets,
$$
N_i=n_i+\half \delta N n_i-\frac{1}{4}\delta N h_i^{~j}n_j
+\frac{1}{4}(-h_i^{~j} + \frac{3}{4}\,h_i^{~l}h_l^{~j})n_j \, .
$$
In \cite{dRGT} the third term comes with a coefficient $1/8$ and the 
fourth term is absent. Thus it appears that at low orders this
relation is not unique.\label{Nn-pert}}, 
\be
N^i = n^i +M^i + N\, D^i_{\,k}\, n^k\,,
\label{NnD}
\ee
where the matrix $D$ is given by (\ref{SolnD}) and (\ref{Minv}) above.
Note that $n^i$ parameterize the difference between the shift functions
of $g_{\mu\nu}$ and $f_{\mu\nu}$. In the following section we will
show that, with this choice of $n^i$, criteria 1 of section
\ref{criteria} is automatically satisfied.

\subsection{The Hamiltonian constraint in the minimal massive theory}

Now we consider the $N$ and $n^i$ equations of motion (\ref{eomNni})
and show that they do not depend on $N$. Hence they give rise to a
Hamiltonian constraint.  Incorporating (\ref{AB}), (\ref{mAmB}),
(\ref{d}) and (\ref{NnD}) into the minimal massive theory
(\ref{LminADM}), leads to an action in terms of the $n^i$ which is
linear in $N$, meeting requirements 2 and 3 of section \ref{criteria},
\bea
{\cal L}_{min}&=&\pi^{ij}\partial_t \gamma_{ij} + N
R^0+R_i \big[n^i+M^i+N\,D^i_{\,k}\,n^k 
 \big] \nn\\[.2cm]
&&\qquad \qquad \qquad \qquad \qquad 
-2m^2\sqrt{\det\gamma}\,\big[\sqrt{x}
+N\sqrt{x}\tr D -3N \big]\,. 
\label{LminNni}
\eea
Thus the $N$ equation of motion (\ref{eomNni}) is independent of $N$
by construction. We now show that the $n^i$ equations are also
independent of $N$ as required by criterion 1. To get the $n^i$
equations, one needs,
\bea
\frac{\p}{\p n^k} \sqrt{x} &=&- \frac{1}{\sqrt{x}} \,n^j f_{ji} \,
\frac{\p n^i }{\p n^k} \, , \\ 
\frac{\p}{\p n^k}(\sqrt{x}\tr D)&=& - \frac{1}{\sqrt{x}} \, 
n^j f_{ji}\,\frac{\p(D^i_{~l}\,n^l  )}{\p n^k} \, .
\eea
The first line easily follows from (\ref{x}). In the second line, we
have first used (\ref{D}) to re-express the left hand side in terms of
the square root matrix and then
$\delta\tr\sqrt{E}=\tfrac{1}{2}\tr(\sqrt{E}^{\,-1}\delta E)$ to
evaluate the derivative. On using (\ref{fD}), the right hand side
follows. Then, varying ${\cal L}_{min}$ in (\ref{LminNni}) with
respect to $n^k$ gives,
\be
\left(R_i+2m^2\sqrt{\det\gamma}\,\frac{n^j f_{ji}}{\sqrt{x}}\right)
\left[\frac{\p}{\p n^k}(n^i+N D^i_{~l} n^l) \right] =0\,. 
\ee
Note that the expression in the square brackets is the Jacobian matrix  
$\frac{\p N^i}{\p n^k}$ of (\ref{NnD}). From (\ref{eomNNi}) it is then 
obvious that the expression in the parenthesis is indeed 
$\p{\cal L}_{min}/\p N^i$. The Jacobian matrix is generically invertible 
as can be checked, for example, perturbatively by using the expression 
in footnote (\ref{Nn-pert}). Hence one gets the $n^i$ (or $N^i$)
equations of motion, 
\be
\sqrt{x} \,R_i+2 m^2\sqrt{\det\gamma}\, f_{ij}\, n^j=0 \, ,
\label{neom-final}
\ee
which are independent of $N$ as advertised. Using (\ref{x}), these
determine $n^i$ in terms of $\gamma_{ij}$ and the conjugate momenta
$\pi^{ij}$ (contained in $R_i$), 
\be
n^i=\frac{-M}{\sqrt{4 m^4{\det\gamma}+R_k (f^{-1})^{kl}R_l}}
  \,\,\,(\tf^{-1})^{ij}\,R_j\,  .
\label{nSoln}
\ee
As a consistency check, note that this implies,
\be
x=\frac{4m^4\det\gamma\,M}{4m^2\det\gamma+R^T\,\tf^{-1} R}\, >0 \, ,
\label{xR}
\ee
so that $\sqrt{x}$ is real on the constraint surface.

The $N$ equation of motion is, 
\be
R^0+R_i D^i_{\,j} n^j -2m^2\sqrt{\det\gamma}\left[
\sqrt{x}\,\, D^k_{~k} -3 \right]=0 \, .
\label{Neom}
\ee
The $n^i$ that appear explicitly and through $x$ (\ref{x}) and $D$
(\ref{SolnD}) can be eliminated using the solution (\ref{nSoln}). Thus
the $N$ equation becomes a constraint on the 12 components of
$\gamma_{ij}$ and $\pi^{ij}$ and reduces the number of canonical
variables to 11. This is the Hamiltonian constraint of the minimal
massive action. 

To see that this constraint has the correct form to eliminate the
ghost, we adapt a parameterization of \cite{ADM} for $\gamma_{ij}$ in
terms of the six functions $\gamma_{ij}^{TT}$(2), $\gamma_{j}^T$(2),
$\gamma^L$(1)1 and $\gamma^T$(1),
\be
\gamma_{ij}=\gamma_{ij}^{TT}+\p_i\gamma_{j}+\p_j\gamma_{i}+
\gamma_{ij}^T \,.
\ee
Here, $\gamma_{ij}^T=\half(\delta_{ij}-\nabla^{-2}\p_i\p_j)\gamma^T$
and $\gamma_i=\gamma_i^T+\p_i \gamma^L$. As the notation implies,
$\gamma_{ij}^{TT}$ is traceless, transverse and $\gamma_i^T$ is
transverse; hence the above counting. $\gamma^T$ is the trace of the
transverse part of $\gamma_{ij}$. The flat space limit indicates that 
$\gamma_{ij}^{TT}$, $\gamma_{j}^T$ and $\gamma^L$ carry the massive
spin-2 graviton while $\gamma^T$ describes the ghost. From the
analysis of \cite{ADM} one can see that $\gamma^T$ appears in
$R^0$ in the right way to be eliminated by (\ref{Neom}), in analogy
with GR. One more constraint is needed to remove the canonically
conjugate variable. 

\subsection{The secondary constraint and the absence of the BD ghost}
 
Now we argue that the Hamiltonian constraint gives rise to a secondary
constraint (for an explicit proof see \cite{HR6}) so that the 12
dimensional phase space of $\gamma_{ij}$ and $\pi^{ij}$ has only 10
degrees of freedom, corresponding to the five polarizations of the
massive graviton.

Before proceeding note that after integrating out the shifts $N^i$,  
the Lagrangian (\ref{LminNni}) remains linear in the lapse $N$,
\be
\label{LC2}
{\cal L}_{min}= \pi^{ij}\partial_t \gamma_{ij} - {\cal H}_0
(\gamma_{ij},\pi^{ij},f)+N\, {\cal C} (\gamma_{ij},\pi^{ij},f) \, . 
\ee
From this one can read off a Hamiltonian, ignoring the usual ADM
contribution that can be expressed as a boundary term,  
$H=\int d^3x( {\cal H}_0- N{\cal C})$.  

A secondary constraint is obtained by demanding that the Hamiltonian
constraint (\ref{Neom}), now summarized as ${\cal C}=0$, is
independent of time on the constraint surface. In the Hamiltonian
formulation this condition is expressed in terms of a Poisson bracket,   
\be
\frac{d}{dt}{\cal C}=\left\{{\cal C},\, H \right\} \approx 0
\ee
with $H$ as given above. Now, if $\{{\cal C}(x),\,{\cal C}(y)\}
\approx \!\!\!\!\!/\,\,0$, this condition becomes an equation for $N$
and does not impose any constraint on the $\gamma_{ij}$ and
$\pi^{ij}$. If this were true, as argued to be the case in
\cite{kluson}, then a second constraint would not exist. However, in
\cite{HR6} it has been shown through explicit calculation that
$\{{\cal C}(x),\, {\cal C}(y)\}\approx 0$. Then the condition $d{\cal
  C}/dt=0$ is independent of $N$ and thus becomes a second constraint
on $\gamma_{ij}$ and $\pi^{ij}$ (with $H_0=\int d^3x\, {\cal H}_0$),   
\be
{\cal C}_{_{(2)}} \equiv \left\{{\cal C}, \, H_0 \right\}\approx 0 \, ,
\ee
provided the expression for ${\cal C}_{_{(2)}}$ does not vanish
identically. That this is the case can be easily shown
perturbatively. By construction, the Lagrangian (\ref{LC2}) reproduces
the Fierz-Pauli Lagrangian at lowest order in the fields and for
$f_{\mu\nu}=\eta_{\mu\nu}$. Hence,  
\be
{\cal H}_0\simeq{\cal H}_0^{FP}+{\cal O}(\gamma^3,\pi^3,\delta f),
~~~~ {\cal C} \simeq{\cal C}^{FP}+{\cal O}(\gamma^2,\pi^2,\delta f)\,. 
\ee 
Therefore, one should find that,
\be
{\cal C}_{_{(2)}} \simeq {\cal C}_{_{(2)}}^{FP} +{\cal
  O}(\gamma^2,\pi^2,\delta f)\, ,
\ee
where it is known that ${\cal C}_{_{(2)}}^{FP}$ is neither
identically zero nor equal to ${\cal C}^{FP}$ (see, for example, 
\cite{Hinterbichler}). Moreover, as can be seen from the Fierz-Pauli
structure, enforcing $d{\cal C}_{_{(2)}}/dt=0$ will result in an
equation for $N$ and no further constraints are generated.
For details see \cite{HR6}. To summarize, this proves the
existence of a secondary constraint that is needed to completely
eliminate the propagating Boulware-Deser ghost mode.   

\subsection{The positivity of the Hamiltonian and open issues}

In general relativity, the Hamiltonian corresponding to the ADM
Lagrangian (\ref{S-GR-ADM}) is expressible as a boundary term
\cite{ADM}. The boundary expression for the Hamiltonian can also be
derived in a more general setup \cite{RT,BY} by considering the 
Gibbons-Hawking boundary terms that have been suppressed in
(\ref{S-GR-ADM}). This boundary expression for the Hamiltonian is
independent of the mass term and remains unchanged in the massive
theory. However, the mass term gives an extra bulk contribution ${\cal 
  H}_m$ to the Hamiltonian due to the reduction in the number of
constraints.

In the massive gravity actions that they analyzed, Boulware and Deser
found that the contribution of the mass term to the Hamiltonian ${\cal
  H}_m$ was generically not positive and moreover, it diverged in the
limit $m\rightarrow 0$ (as reviewed in the appendix). The pathological
behavior of the Hamiltonian was given as a reason for disregarding
massive gravity \cite{BD,BD2}. 

Here we consider the contribution of the mass term to the Hamiltonian
in the minimal massive action with a general reference metric. This
can be easily computed from ${\cal L}_{min}$ in (\ref{LminNni}) upon
imposing the Hamiltonian constraint (\ref{Neom}). Using (\ref{nSoln})
and (\ref{xR}) this contribution becomes,
\be
{\cal H}_{min,m}=M\sqrt{4m^4\det\gamma+R_k\,(\tf^{-1})^{kl} R_l}\, 
-\,R_i\,M^i \, ,
\label{Hm}
\ee 
where, $M$ and $M^i$ are the lapse and shift functions of $f_{\mu\nu}$.
Note that for $f_{\mu\nu}=\eta_{\mu\nu}$ ($M=1, M^i=0$), which was the
case considered by Boulware and Deser, this is clearly positive and
well behaved in the limit $m\rightarrow 0$, avoiding the pathologies 
observed in \cite{BD}. The same applies to any reference metric
$f_{\mu\nu}$ for which $M^i=0$ and $M>0$. 

In general, when $M^i\neq 0$, the last term in ${\cal H}_{min,m}$
appears problematic, but at least in some cases this is a gauge
artifact. For example, even for a flat $f_{\mu\nu}=\p_\mu\phi^a
\p_\nu\phi^b\eta_{ab}$, if we do not choose the physical gauge
$\phi^\mu=x^\mu$, the Hamiltonian is not manifestly positive. But
clearly, in this case, the problem is a gauge artifact. It seems that
in all situations where one can choose a gauge with $M^i=0$, the
Hamiltonian remains positive.

However, while the minimal massive action with a general reference
metric is free of Boulware-Deser instability, it seems that it may not
always have a positive Hamiltonian. In fact, given the known
instabilities of massive gravity in, say, de Sitter and anti de Sitter
backgrounds, we do not expect a massive gravity Hamiltonian to be
positive for all possible $f_{\mu\nu}$ in all regions of parameter
space. However, precisely because of the $R_iM^i$ structure, the
theory is potentially linear in the lapse and shift functions of
$f_{\mu\nu}$, provided the constraints do not introduce a non-linear
dependence through the elimination of the ghost mode. This opens up
the possibility of consistently promoting $f_{\mu\nu}$ to a dynamical
variable. This will be discussed in detail in \cite{HR5}.

\subsection{An alternative set of variables}\label{hats}

It is not apparent that the equations of section 3.1 depend in a
simple way on the lapse $M$ of $f_{\mu\nu}$. However, the solution
(\ref{nSoln}) shows that the $n^i$  are linear in $M$, so that $x$
is proportional to $M^2$ and the matrix $D$ goes as $1/M$. This
motivates working with an alternative set of variables, 
\be
n^i=M\hat n^i\,,\qquad D^i_{\,j}=\hat D^i_{\,j}/M.
\ee
which makes the simple dependence of the theory on $M$ manifest even
beyond the minimal model. In term of these, (\ref{NnD}) takes the
form, 
\be
N^i = M\hat n^i +M^i + N\, \hat D^i_{\,k}\, \hat n^k\,,
\label{NnDhat}
\ee
and 
\be
x=M^2 \hat x\,,\qquad \hat x= 1-\hat n^T\tf \hat n
\ee
The defining equation for $\hat D$, and hence the matrix $\hat D$
itself, is independent of $M$,
\be
\sqrt{\hat x}\,\hat D =\sqrt{(\gamma^{-1}-\hat D \hat n \hat n^T\hat
  D^T)\,\tf} \, , 
\label{Dhat}
\ee 
Then it follows that, expressed in terms of hatted variables, the
matrix $\mB$ in (\ref{mAmB}) is independent of $M$. The matrix $\mA$
depends on $M$ in a more complicated way through $a_0=M^2+M\hat
n^lM_l$, $a_i=M\hat n^l M_l$ and $c^i=M\hat n^i+M^i$. However, the
most general massive action (\ref{act2}) contains $\mA$ only in the
combinations $\tr(\mA)$, $\tr(\mA\mB)$ and $\tr(\mA\mB^2)$ (see the
next section). It is easy to verify that all these are linear in $M$.  
Thus, on using the hatted variables, the action (\ref{act2}) becomes
linear in both $N$ (see below) and $M$. 

\section{Absence of the BD ghost in the complete 2-parameter massive
  action}   

We now demonstrate the absence of the Boulware-Deser ghost in the full
two-parameter generalization of the minimal massive theory with a
general reference metric $f_{\mu\nu}$. We show that the Hamiltonian
constraint is maintained even in this case. The analysis is more
involved but it turns out that the variables identified in the minimal
massive model can be used in these more general cases without any
modifications. The argument for the existence of the secondary
constraint parallels the discussion for the minimal model. The details
are given in \cite{HR6}. 

\subsection{The 2-parameter action in terms of the new variables}

We now show that, using the same $n^i$ of the previous section
(\ref{NnD}), the general massive action (\ref{act2}) turns out to be
linear in $N$ thereby satisfying criteria 2 and 3 of section
\ref{criteria}.  In the ADM parameterization, the general Lagrangian is
given by
\be
\label{Lgen}
{\cal L}=\pi^{ij}\partial_t \gamma_{ij} + N R^0 +N^i R_i 
+2m^2\sqrt{\gamma}\,N\,\left(\sum_{n=0}^3\beta_n e_n(\sqrt{g^{-1}f}\,)
\right) \, .
\ee
To check if this Lagrangian is linear in the lapse $N$, let us express
the potential in terms of the matrices $\mA$ and $\mB$ of (\ref{AB}).
Since $e_n\sim(\sqrt{g^{-1}f}\,)^n$, it might seem that negative
powers of $N$ should appear in the Lagrangian.  However, due to the
property $\tr(\mA^k) = (\tr \mA)^k$, these terms cancel amongst
themselves. The potential terms (\ref{e_n}) then give,
\be
\label{Nes}
\begin{array}{lcl}
N e_0 (\sqrt{g^{-1}f} \,) &=& N \, , \\[.2cm]
N e_1 (\sqrt{g^{-1}f} \,) &=& \tr \mA + N \tr \mB \, , \\[.2cm]
N e_2 (\sqrt{g^{-1}f} \,) &=& \tr \mA \tr \mB-\tr \mA \mB+ 
\tfrac{1}{2}N\left[ (\tr \mB)^2-\tr\mB^2 \right] \, , \\[.2cm]
N e_3 (\sqrt{g^{-1}f} \,) &=& \tr \mA \mB^2-\tr \mA \mB \tr \mB +
\tfrac{1}{2} \tr \mA \left[ (\tr \mB)^2-\tr\mB^2  \right]  \\[.2cm]
&&+ \tfrac{1}{6}N\left[ (\tr \mB)^3-3\tr \mB \tr\mB^2 +2\tr \mB^3 \right]\,.
\end{array}
\ee
These terms are at most linear in the lapse $N$ and thus satisfy
criterion 2. The $e_0$ term just contributes to a cosmological
constant while the $e_1$ term was already considered in the previous
section for the minimal action.  So the terms that remain to be
investigated are $e_2$ and $e_3$.

\subsection{The Hamiltonian constraint in the 2-parameter theory} 

We now compute the $n^i$ equations of motion and show that they do not
depend on the lapse $N$. The $R_iN^i$ term in the action contributes a
term $R_i\,J^i_{~k}$ to the equations of motion, where,
\be
J^i_{~k}\equiv\frac{\p N^i}{\p n^k}=\frac{\p}{\p n^k}(n^i+ND^i_{~j}n^j) \, ,
\label{J}
\ee
is the Jacobian matrix of (\ref{NnD}).  Since the Jacobian contains
the lapse $N$, the only way for the $n^i$ equations of motion to be
independent of $N$ is if the contribution of the potential terms is
also proportional to the Jacobian.  This must happen separately for
the $e_2$ and $e_3$ terms.  We show now that this is the case.
  
In the following, we employ matrix notation where $n$ is a column
vector with elements $n^i$ and $n^T$ is its transpose.  To vary with
respect to $n^k$, consider the $\mA$ terms first.  Using (\ref{mAmB}),
these terms are, 
\be
\label{ABs}
\begin{array}{lcl}
\tr \mA &=& \sqrt{x} \, , \\[.2cm]
\tr \mA\mB &=&  -n^T\,\tf\,D\,n \, ,  \\[.2cm]
\tr \mA\mB^2 &=&  - \sqrt{x} \, n^T\,\tf\,D^2\,n  
= -\frac{1}{\sqrt{x}} \, n^T\,\tf\,(\gamma^{-1}-Dnn^TD^T)\,\tf\,n\,.
\end{array}
\ee
The last expression is written in two equivalent ways using
(\ref{D}). Varying these gives  
\be
\label{dABS}
\begin{array}{lcl}
\frac{\p}{\p n^k} \tr \mA &=& - \frac{1}{\sqrt{x}}\, n^T\,
\tf\, \frac{\p n}{\p n^k} \, , 
\\[.2cm] 
\frac{\p}{\p n^k}\tr \mA\mB &=&- n^T\,\tf\,D\,\frac{\p n}{\p n^k}   
\, - \, n^T\,\tf\,\frac{\p(D\,n)}{\p n^k}\,,
\\[.2cm]
\frac{\p}{\p n^k}\tr\mA\mB^2&=&\frac{1}{\sqrt{x}}\, 
(n^T\,\tf D^2\,n)\, n^T\,\tf\,\frac{\p n}{\p n^k}
- 2\sqrt{x}\,n^T\,\tf\,D\,\frac{\p (D\,n)}{\p n^k}
\\[.2cm]
&=&-\frac{1}{\sqrt{x}}\,(n^T\,\tf D^2n)\,n^T\,\tf\,\frac{\p n}{\p n^k}   
- 2\sqrt{x}\,n^T\,\tf D^2\,\frac{\p n}{\p n^k}
\\[.2cm]
&&+\, \frac{2}{\sqrt{x}}\,(n^T\,\tf Dn)\, n^T\,\tf\,
\frac{\p(D\,n)}{\p n^k} \, .
\end{array}
\ee
From the last equality we derive the relation,
\be
\label{simp}
\left[\sqrt{x}\, D + \frac{1}{\sqrt{x}}\,Dn n^T\,\tf\right]  
\frac{\p(D\,n)}{\p n^k} 
 =\left[ \sqrt{x}\,D^2+\frac{1}{\sqrt{x}}\,\one (n^T\,\tf D^2 n)\right]
 \frac{\p n }{\p n^k} \, .
\ee
This relation is very useful. We need all derivatives in $\p(Ne_3)/\p
n^k$ to appear in the combination (\ref{J}). However on direct
substitution we will find some $\p(Dn)/\p n^k$ terms without 
the factor $N$. Equation (\ref{simp}) allows us to re-express these in
terms of $\p (n)/\p n^k$.

Now consider the $\mB$ terms. Using (\ref{mAmB}) these are,
\be
\label{Bs}
\tr \mB=\sqrt{x} \, \tr D \, , \qquad\quad 
\tr \mB^2 = \sqrt{x}^{\,2} \tr \,D^2 \, , \qquad\quad
\tr \mB^3 =  \sqrt{x}^{\,3} \, \tr D^3 \, .
\ee
The variations can be written as,
\be
\label{dBs}
\begin{array}{lcl}
\frac{\p}{\p n^k}\tr\mB&=&-\frac{1}{\sqrt{x}}\,n^T\,\tf\,\frac{\p(Dn)}
     {\p n^k}\, , \\[.2cm] 
\frac{\p}{\p n^k}\tr\mB^2&=&-2\,n^T\,\tf\,D\,\frac{\p(Dn)}{\p n^k}\,,
  \\[.2cm] 
\frac{\p}{\p n^k}\tr\mB^3&=&-3\sqrt{x}\,n^T\,\tf\,D^2\,\frac{\p(Dn)} 
{\p n^k}\, .
\end{array}
\ee
Combining all these results gives
\bea
\tfrac{\p}{\p n^k} N e_1(\sqrt{g^{-1}f})&=&-\frac{1}{\sqrt{x}}
\,n^T\,\tf\, \tfrac{\p}{\p n^k}(n+N D n)\, , 
\nn\\[.1cm]
\tfrac{\p}{\p n^k}Ne_2(\sqrt{g^{-1}f})&=&n^T\,\tf\,
\left[D-\one\tr D\right] \,\tfrac{\p}{\p n^k}(n+N D n) \,,
\label{dNes}\\[.1cm]
\tfrac{\p}{\p n^k}Ne_3(\sqrt{g^{-1}f})&=&-\sqrt{x}\, n^T\,\tf\, 
\left[D^2-D\tr D+\tfrac{1}{2}\one\left[(\tr D)^2-\tr(D^2)\right]
\right]\tfrac{\p}{\p n^k}(n+N D n)\,.\nn
\eea
where (\ref{simp}) was used to simplify the last expression.
Note that the right hand sides are proportional to $J^i_{~k} = \p
N^i/\p n^k$ (\ref{J}) which was a requirement for the $n^i$ equations
of motion to be independent of $N$. So, finally, varying the general
action (\ref{Lgen}) with respect to $n^i$ gives the $N$-independent
equations of motion,
\begin{multline}
\label{eomgen}
R_i-2m^2\sqrt{\gamma}\,\frac{n^lf_{lj}}{\sqrt{x}}\bigg[\beta_1\,
\delta^j_{i} +\beta_2\,\sqrt{x}\, (\delta^j_{i} D^m_{~m}-D^j_{~i})\\
+\beta_3 \, \sqrt{x}^{\,2} \, \left\{\tfrac{1}{2}\delta^j_{~i} 
(D^m_{~m} D^n_{~n}\, -\,D^m_{~n} D^n_{~m}) +D^j_{~m} D^m_{~i}\, -\,
D^j_{~i} D^m_{~m}\right\} \bigg] = 0 \, .
\end{multline}
In principle, these equations can be solved to determine the $n^i$ in
terms of $R_i$ at least perturbatively, although unlike the minimal
massive model, an explicit non-linear solution may be difficult to
obtain. 

The $N$ equation of motion is,
\begin{multline}
R^0+R_i D^i_{\,j}n^j
+2m^2\sqrt{\gamma}\Big[\beta_0+\beta_1 \sqrt{x} \,D^i_{~i}
+\tfrac{1}{2}\beta_2\sqrt{x}^{\,2}\,(D^i_{~i}D^j_{~j}-D^i_{~j}D^j_{~i})
\\
+\tfrac{1}{6}\beta_3\sqrt{x}^{\,3}\,(D^i_{~i}D^j_{~j}D^k_{~k}-
3D^i_{~i}D^j_{~k} D^k_{~j}+2D^i_{~j} D^j_{~k} D^k_{~i} ) \Big]=0 \, . 
\end{multline}
Eliminating the $n^i$ in favour of $R^i$ converts this into the 
Hamiltonian constraint on $\gamma_{ij}$ and $\pi^{ij}$. This and 
its associated secondary constraint are enough to eliminate the 
ghost. The existence of a non-trivial secondary constraint follows
from an argument similar to one for the minimal model in the previous
section. For the explicit computations see \cite{HR6}.

This demonstrates the existence of a two-parameter family of
non-linear theories of massive gravity with general $f_{\mu\nu}$ that
do not suffer from the Boulware-Deser ghost instability.

\section{Discussions}

In this work we have shown that the recently proposed non-linear
massive gravity theories do not suffer from the Boulware-Deser ghost
instability for an arbitrary non-dynamical reference metric.  This is
a generalization of the work in \cite{HR3} which showed the absence of
the BD ghost for a flat reference metric.  To reiterate, the
appearance of the BD ghost is due to the absence of a Hamiltonian
constraint. We have shown that the massive actions (\ref{act2})
contain such a constraint and an associated secondary constraint and
hence are free from this instability.

The theory discussed here need not necessarily be interpreted as a
theory of massive gravity, which may or maynot be consistent with
observations. It also has an alternative interpretation as a ghost
free theory of a massive spin-2 field $g_{\mu\nu}$ (say, a meson) in a
fixed gravitational background given by $f_{\mu\nu}$.

Much of the recent analysis of the ghost issue in massive gravity has
relied on the St\"uckelberg formulation and a flat $f_{\mu\nu}$.  In
the decoupling limit, this formulation provides a powerful tool for
studying the ghost content of the theory because of the
Goldstone-vector boson equivalence theorem.  The recently proposed
massive gravity actions, which we have shown to be ghost-free even for
a general $f_{\mu\nu}$, were constructed by demanding the absence of
the BD ghost in the St\"uckelberg formulation and in the decoupling
limit alone \cite{dRG,dRGT}.

However, away from the decoupling limit, the equivalence theorem is no
longer valid and, even for a flat $f_{\mu\nu}$, the ghost analysis in
the St\"uckelberg formulation becomes significantly more involved.  A
full analysis of the constraints and gauge conditions of the theory
must be performed to obtain the physical spectrum in order to identify
the ghost. Some recent work which does not take these issues into
account has suggested that, in the St\"uckelberg formulation (or
relatedly, using a helicity decomposition), the BD ghost inevitably
reappears away from the decoupling limit, at higher orders in
perturbation theory \cite{ACM,CM,FPW}. These results are in 
contradiction with the conclusions of \cite{HR3} and its
generalization in the present work. However, a thorough analysis
of the ghost issue in the perturbative St\"uckelberg framework
was performed in \cite{dRGT3,dRGT4}. This showed that when the
constraints are taken into account, the BD ghost is indeed absent and
the apparent discrepancy between the perturbative St\"uckelberg
approach and the non-linear analysis of \cite{HR3} disappears.

By generalizing the ghost analysis to general $f_{\mu\nu}$, the
results of this paper open up the possibility of finding new and
interesting classical solutions to massive gravity theories. Moreover,
it is known at the linear level that massive gravity in FRW-type
backgrounds may contain instabilities that are distinct from the
Boulware-Deser ghost. As argued in \cite{Grisa,BDH1,BDH2}, these
problems might be avoided in the full non-linear theory in a dynamical
process. The actions studied here provide a non-linear setup in which
this issue might be investigated. Finally, the results of this paper
provide a first step towards promoting $f_{\mu\nu}$ to a dynamical
variable and thus creating a consistent bimetric theory of gravity
(see, \cite{HR5}).

An interesting possibility is the realization of ghost free massive
gravity within string theory setups. While this is not obvious at the
level of fundamental string, there is evidence that linear massive
gravity in AdS background arises within the AdS/CFT framework 
\cite{Kiritsis:2006hy,Aharony:2006hz}. It is interesting to check if
AdS/CFT could also reproduce the correct non-linear generalizations
described in this paper.   

\vskip.3cm

\noindent {\bf Acknowledgements:} We would like to thank F. Berkhahn,
S. Deser, J. Enander, S. Hofmann, J. Kluson, B. Sundborg, and M. von
Strauss for discussions and comments on the draft.  We are especially
grateful to F. Berkhahn for comments and suggestions on the
presentation of section 2.4. The majority of this work was completed
while R.A.R. was supported by the Swedish Research Council (VR)
through the Oskar Klein Centre.  R.A.R. is currently supported by NASA
under contract NNX10AH14G.

\appendix

\section{Appendix: Further review of the ghost problem}
\subsection{Absence of ghost in general relativity}

In field theory, a ghost refers to a physical mode with negative
kinetic energy. In the quantum theory this results in states with
negative probability. When the action is not in diagonal form in the
fields and particularly in the presence of constraints and gauge
symmetries, the physical content of the theory may not be directly
discernible. To identify the ghost in such cases, one has to first
determine the physical degrees of freedom with canonical kinetic
terms. Alternatively, the ghost appears in the 2-point function as a
mass pole with negative residue\footnote{For fields $\phi_a$ with
  propagators $G_{ab}$ interacting with external sources $J^a$, the
  transition amplitude $\langle0,out|0,in\rangle_J \sim e^{i\int
    J^aG_{ab} J^b}$ will be less than 1 if $J^a$ excite $|0,in\rangle
  $ into particle states of positive probability.  But this could
  exceed unity in the presence negative probability states, implying a
  positive exponent, $i(2\pi i)Res(JGJ,k^0=|\vec k|)>0$,
  \cite{BD}.}. This comes handy when propagators are known.

That metric fluctuations in a modified theory of gravity could easily
contain a ghost can be inferred by investigating linearized general 
relativity. Decomposing the metric fluctuations $h_{\mu\nu}=g_{\mu\nu}  
-\eta_{\mu\nu}$ in terms of its traceless transverse ($h^\perp_{\mu\nu}$),
transverse vector ($a^\perp_\mu$), longitudinal ($\phi$) and scalar
($s$) parts,  
\be
h_{\mu\nu}=h^\perp_{\mu\nu}+\p_{(\mu} a^\perp_{\nu)}+\p_\mu\p_\nu\phi
+\tfrac{1}{4}\eta_{\mu\nu} s
\label{h-decomp1}
\ee
one can easily check that the quadratic Einstein-Hilbert action
depends only on the five components of $h^\perp$ and the sixth scalar
mode $s$, 
\be
S_{EH}= \frac{1}{4}M_p^2\int d^4x \Big [h^{\perp\mu\nu}\,\Box\,
h^\perp_{\mu\nu}-\frac{3}{8}s\,\Box\, s \Big]\,.
\ee 
The other modes drop out due to invariance under $\delta
h_{\mu\nu}=-\p_{(\mu} \xi_{\nu)}$. Obviously $s$ has a negative
kinetic term and is potentially a ghost, but in general relativity
it does not survive as a propagating mode. In more detail, $s$ 
contributes to the graviton propagator $i\Delta^{(0)}$ as a ghost.
Indeed it contributes the second term in the saturated $k$-space
propagator, 
\be
T^{\mu\nu} \Delta^{(0)}_{\mu\nu\rho\sigma}T^{*\rho\sigma} 
=\frac{1}{k^2} \Big[ 
T^{\perp\mu\nu}\,{T}_{\mu\nu}^{*\perp} -\frac{1}{6}\,T\, T^* \Big]\,
\label{Gm=0}
\ee
which obviously has a negative residue at the $k^2=0$ pole. However,
for $k^2=0$, {\it and only for this value}, the ghost is
cancelled by a similar contribution from the first term, coming from
the helicity zero component of the spin-2 field $h^\perp$. This is
discussed below. The cancellation is peculiar to the massless
theory and does not takes place otherwise.

\subsection{Avoiding the ghost in Fierz-Pauli massive gravity}
   
Metric fluctuations around flat spacetime are made massive by adding
the Fierz-Pauli mass term to the linearized EH action
\cite{FP1,FP2,vDV},  
\be
S_{FP}=S_{EH}[h] -\frac{1}{4}M_p^2 m^2 \int d^4x \left[
h^\mu_\nu\, h_\mu^\nu-a \, (h^\mu_\mu)^2\,\right] \,.
\label{FP}
\ee
where, $h^\mu_\nu=\eta^{\mu\rho} h_{\rho\nu}$. In this form, the mass
term is not gauge invariant and depends on the $a_\mu$ and $\phi$
of (\ref{h-decomp1}). It modifies the massless propagator 
(\ref{Gm=0}) by  shifting the mass poles to $k^2=-m^2$ and
$k^2=-m^2_s=\frac{1}{2} \frac{1-4a}{1-a}\,m^2$. Then, on using $k_\mu
T^{\mu\nu}=0$ (see for example, \cite{HHV}),
\be
T^{\mu\nu} \Delta^{(m)}_{\mu\nu\rho\sigma}T^{*\rho\sigma} 
=\Big[ \frac{1}{k^2+m^2}\,T^{\perp\mu\nu}\,{T}_{\mu\nu}^{*\perp} 
-\frac{1}{6}\frac{1}{k^2+m_s^2}\,T\, T^* \Big]\,.
\label{Gm}
\ee
Then, as shown below, there is no way of cancelling the wrong sign
contribution from the scalar part against the spin-2 part and all 6
modes (including the ghost) contribute to the propagation. The only
way out is to tune $a=1$ so that $m^2_s=\infty$.  This decouples the
ghost keeping only the 5 healthy polarizations of the massive
graviton.

It was pointed out by Boulware and Deser \cite{BD} that this method of
avoiding the ghost cannot be easily implemented beyond the linear
theory. Then, generically the theory will contain 6 propagating modes 
indicating the reappearance of the $6th$ mode that was removed at the
linear level by setting $a=1$. We discuss this below.

\subsection{Unitarity analysis of the saturated propagator}

The treatment here follows \cite{Nunes}. In our conventions, a
negative residue of the saturated propagator
$T^{\mu\nu}\Delta_{\mu\nu\rho\sigma}T^{*\rho\sigma}$ implies the 
presence of a ghost. Given a 4-vector $k^\mu$ we can construct a set
of 4 linearly independent vectors,
\be
k^\mu=(k^0,\vec k )\,,\qquad \wt k^\mu=(-k^0,\vec k)\,, \qquad
\epsilon^\mu_r= (0, \vec\epsilon_r)\,\quad \mbox{\rm for}\quad (r=1,2)
\ee
such that $\vec\epsilon_r\cdot\vec\epsilon_s=\delta_{rs}$ and $\vec k
\cdot\vec\epsilon_r=0$. Note that $(k\wt k)\equiv k_\mu\wt
k^\mu=(k^0)^2+|\vec k|^2 >0$ and $k^2=\wt k^2$. In this basis, a
generic symmetric tensor can be expanded as  
\bea
T_{\mu\nu}(k)&=&ak^\mu k^\nu+b\wt k^\mu\wt k^\nu+\tfrac{1}{2}
c^{rs}(\epsilon^\mu_r \epsilon^\nu_s+\epsilon^\nu_r \epsilon^\mu_s)
+\tfrac{1}{2} d(k^\mu\wt k^\nu + k^\nu\wt k^\mu)
+\tfrac{1}{2} e^r(k^\mu\epsilon^\nu_r+k^\nu\epsilon^\mu_r) \nn\\
&&\hfill +\tfrac{1}{2}f^r(\wt k^\mu\epsilon^\nu_r+
\wt k^\nu\epsilon^\mu_r)   
\eea
The conservation equation $k^\mu T_{\mu\nu}=0$ implies $b={a
  k^4}/{(k\wt k)^2}$, $d=-{2a k^2}/{(k\wt k)}$ as well as $f^r=
-{k^2}/{(k\wt k)}e^r$. 
The propagator (\ref{Gm=0}) in Einstein-Hilbert theory takes the
standard from on using $T_{\mu\nu}^\perp=T_{\mu\nu}- 
\tfrac{1}{3}(\eta_{\mu\nu}-\frac{k^\mu k^\nu}{k^2})T$. Then in the
above parameterization of $T_{\mu\nu}$ the residue at the zero mass
pole is positive, hence the theory is ghost free,
\be
\Big[T^{\mu\nu}T^*_{\mu\nu}-\tfrac{1}{2}TT^*\Big]_{k^2=0}=\tfrac{1}{2}
|c_{11}-c_{22}|^2 + 2|c_{12}|^2 > 0 \,.
\ee
In the massive theory, the propagator has the generic momentum space
form (\ref{Gm}). The $T^\perp T^{\perp *}$ term is due to spin-2
exchange and gives a positive residue at the mass pole, 
\bea
\Big[T^{\mu\nu}T^*_{\mu\nu}-\tfrac{1}{3}TT^*\Big]_{k^2=-m^2}&=&
\Big[ \tfrac{2}{3}\Big|a k^2 (1-\tfrac{k^2 \wt k^2}{(k\wt k)^2})
+\frac{c}{2}\Big|^2+\tfrac{1}{2}|c_{11}-c_{22}|^2 + 2|c_{12}|^2
\nn\\[.1cm]
&&\qquad +\tfrac{1}{2}(|e^1|^2+|e^2|^2)\,k^2
(\tfrac{\wt k^2}{(k\wt k)}-1) \Big]_{k^2=-m^2}>0 
\label{spin2}
\eea
The first line is manifestly positive. In the second line,
$k^2(\frac{\wt k^2}{(k\wt k)}- 1) |_{k^2=-m^2} =
m^2(\frac{m^2}{m^2+2|\vec k|^2}+1)>0$, hence the overall positivity.  

The $T T^*$ term in (\ref{Gm}) is due to the exchange of the scalar
mode $s$ of mass $m_s$ which is a ghost for any finite mass since,
\be
-\frac{1}{6}TT^*\Big|_{k^2=-m_s^2}= -\tfrac{1}{6} 
\Big|a k^2(1-\tfrac{k^2 \wt k^2}{(k\wt k)^2})-c\Big|^2\Big|_{k^2=-m_s^2} 
< 0
\ee
Only for $k^2=0$ this cancels against a contribution in the $T^\perp
T^\perp$ term resulting in the healthy GR expression above. The only other
possibility to get rid of this ghost is to take $m_s\rightarrow\infty$,
for fixed $m$. This decouples the scalar ghost and retains only the
healthy spin-2 contribution (\ref{spin2}). This is the Fierz Pauli
massive gravity for $a=1$. 

\subsection{The Boulware-Deser analysis}

As a specific example, \cite{BD} considers a FP-type mass
(\ref{FP-bg}) where $h_{\mu\nu}=g_{\mu\nu}-\eta_{\mu\nu}$ is
not treated as a small perturbation. Then, in ADM variables, 
\be
h^\mu_\nu\, h_\mu^\nu- \, (h^\mu_\mu)^2=
(h_{ij})^2-(h)^2+2 h(1-N^2+N_iN_j \gamma^{ij})-2N_iN_i
\label{FP-nl}
\ee
where $h_{ij}=\gamma_{ij}-\delta_{ij}$ and $h=h_{ii}$. This is
obviously non-linear in both $N$ and $N^i$. The equations of motion
for these can be solved to give,
\be
N=-\frac{R^0}{m^2\,h}\,,\qquad 
N_i=\frac{1}{m^2}\left[(h\gamma^{-1}-\I^{-1})^{-1}\right]_{ij}R^j 
\ee
There is no constraint on the remaining variables and the theory
contains 6 propagating modes, including the ghost. Substituting back
in the action one can compute the Hamiltonian density ${\cal H}=
\pi^{ij}\p_t\gamma_{ij}-{\cal L}$ (ignoring the boundary contribution
that is the same as in GR),
\be
{\cal H}=\frac{m^2}{4}\left[(h_{ij})^2-h^2\right]+\half h m^2
+\half\frac{(R^0)^2}{m^2 h}-\half\frac{1}{m^2}
R^i\left[(h\gamma^{-1}-\I^{-1})^{-1}\right]_{ij}R^j 
\ee
\cite{BD} noted that the corresponding Hamiltonian is not always 
positive and that it diverges in the limit $m\rightarrow 0$. Hence the
conclusion that gravity cannot have a finite range.

It is instructive to see how the ghost disappears in the linear FP
limit. In this case, expanding around a flat background, $\delta
N=N-1$, $\delta N_i=N_i$ and $h_{ij}=\gamma_{ij}-\delta_{ij}$ are
small perturbations and to {\it quadratic order} the FP mass term is
linear in $\delta N$,
\be
h^\mu_\nu\, h_\mu^\nu- \, (h^\mu_\mu)^2=
(h_{ij})^2-(h)^2 -4 h\delta N -2\delta N_i\delta N_i
\ee 
The $N_i$ equations $R^i(h,\delta\pi)=m^2\,\delta N^i$
determine the $\delta N_i$. But the $\delta N$ equation
$R^0(h,\delta\pi)=m^2\,h_{ii}$ is independent of lapse and shift and
becomes the modified Hamiltonian constraint. The requirement that this
constraint is maintained under time evolution, results in a
non-trivial secondary constraint on the $(h_{ij}, \delta\pi^{ij})$.
These 2 constraints reduce the number of independent ($h,\delta\pi$)
components from 12 to 10, implying the existence of only 5 propagating
modes and no ghost mode, consistent with the manifestly covariant
analysis of the propagator.


\begin{thebibliography}{999}  
\bibitem{BD}  
  D.~G.~Boulware and S.~Deser,  
  Phys.\ Rev.\  D {\bf 6}, 3368 (1972).  
\bibitem{BD2}  
  D.~G.~Boulware, S.~Deser,
  Phys.\ Lett.\  {\bf B40 } (1972)  227-229.
\bibitem{dRG}   
  C.~de Rham and G.~Gabadadze,   
  Phys.\ Rev.\  D {\bf 82}, 044020 (2010)   
  [arXiv:1007.0443 [hep-th]].   
\bibitem{dRGT}   
  C.~de Rham, G.~Gabadadze and A.~J.~Tolley,   
  arXiv:1011.1232 [hep-th].   
\bibitem{HR3}
S.~F.~Hassan and R.~A.~Rosen,
  arXiv:1106.3344 [hep-th].
\bibitem{HR1}
  S.~F.~Hassan and R.~A.~Rosen,
  arXiv:1103.6055 [hep-th].
\bibitem{dRGT2}
C.~de Rham, G.~Gabadadze and A.~J.~Tolley,
  arXiv:1107.0710 [hep-th].
\bibitem{dRGT3}
 C.~de Rham, G.~Gabadadze and A.~Tolley,
  arXiv:1107.3820 [hep-th].
\bibitem{dRGT4}
  C.~de Rham, G.~Gabadadze, A.~J.~Tolley,
  [arXiv:1108.4521 [hep-th]].
\bibitem{D'Amico}
 G.~D'Amico, C.~de Rham, S.~Dubovsky, G.~Gabadadze, D.~Pirtskhalava and
 A.~J.~Tolley, 
 arXiv:1108.5231 [hep-th].
\bibitem{ISS}
  C.~J.~Isham, A.~Salam and J.~A.~Strathdee,
  Phys.\ Rev.\  D {\bf 3}, 867 (1971).
\bibitem{SS}
  A.~Salam and J.~A.~Strathdee,
  Phys.\ Rev.\  D {\bf 16}, 2668 (1977).
\bibitem{FP1}  
  M.~Fierz,  
  Helv.\ Phys.\ Acta {\bf 12} (1939) 3.  
\bibitem{FP2}  
  M.~Fierz and W.~Pauli,  
  Proc.\ Roy.\ Soc.\ Lond.\  A {\bf 173} (1939) 211.  
\bibitem{Higuchi}
  A.~Higuchi,
  Nucl.\ Phys.\  B {\bf 282} (1987) 397.
\bibitem{DW}
  S.~Deser and A.~Waldron,
  Phys.\ Lett.\  B {\bf 508} (2001) 347
  [arXiv:hep-th/0103255].
\bibitem{Porrati}   
M.~Porrati,   
  Phys.\ Lett.\  B {\bf 498}, 92 (2001)   
  [arXiv:hep-th/0011152].   
\bibitem{KMP}   
  I.~I.~Kogan, S.~Mouslopoulos and A.~Papazoglou,   
  Phys.\ Lett.\  B {\bf 503}, 173 (2001)   
  [arXiv:hep-th/0011138].   
\bibitem{Grisa}
  L.~Grisa and L.~Sorbo,
  Phys.\ Lett.\  B {\bf 686} (2010) 273
  [arXiv:0905.3391 [hep-th]].
\bibitem{BDH1}
  F.~Berkhahn, D.~D.~Dietrich and S.~Hofmann,
  JCAP {\bf 1011} (2010) 018
  [arXiv:1008.0644 [hep-th]].
\bibitem{BDH2}
  F.~Berkhahn, D.~D.~Dietrich and S.~Hofmann,
  arXiv:1104.2534 [hep-th].
 \bibitem{ADM}
  R.~L.~Arnowitt, S.~Deser and C.~W.~Misner,
  arXiv:gr-qc/0405109.
\bibitem{HR6}
  S.~F.~Hassan and R.~A.~Rosen,
  arXiv:1111.2070 [hep-th].
\bibitem{BDZ1}
  E.~Babichev, C.~Deffayet and R.~Ziour,
  Phys.\ Rev.\  D {\bf 82} (2010) 104008
  [arXiv:1007.4506 [gr-qc]].
\bibitem{TMN} 
T.~M.~Nieuwenhuizen, 
  arXiv:1103.5912 [gr-qc]. 
\bibitem{Koyama1}
  K.~Koyama, G.~Niz and G.~Tasinato,
  arXiv:1103.4708 [hep-th].
\bibitem{Koyama2}
  K.~Koyama, G.~Niz and G.~Tasinato,
  arXiv:1104.2143 [hep-th].
 \bibitem{AGS}  
  N.~Arkani-Hamed, H.~Georgi and M.~D.~Schwartz,  
  Annals Phys.\  {\bf 305} (2003) 96  
  [arXiv:hep-th/0210184].  
\bibitem{HR5}
  S.~F.~Hassan and R.~A.~Rosen,
  arXiv:1109.3515 [hep-th].
\bibitem{Hinterbichler}
  K.~Hinterbichler,
  arXiv:1105.3735 [hep-th].
\bibitem{HRS2}
S.~F.~Hassan and A. Schmidt-May, In preparation.
\bibitem{CNPT}  
  P.~Creminelli, A.~Nicolis, M.~Papucci and E.~Trincherini,  
  JHEP {\bf 0509}, 003 (2005)  
  [arXiv:hep-th/0505147].  
\bibitem{kluson}  
  J.~Kluson,
  arXiv:1109.3052 [hep-th].
\bibitem{RT}
  T.~Regge and C.~Teitelboim,
  Annals Phys.\  {\bf 88} (1974) 286.
\bibitem{BY}
  J.~D.~Brown and J.~W.~York,
  Phys.\ Rev.\  D {\bf 47} (1993) 1407
  [arXiv:gr-qc/9209012].
\bibitem{ACM} 
  L.~Alberte, A.~H.~Chamseddine and V.~Mukhanov, 
  JHEP {\bf 1104}, 004 (2011) 
  [arXiv:1011.0183 [hep-th]]. 
\bibitem{CM}
  A.~H.~Chamseddine and V.~Mukhanov,
  arXiv:1106.5868 [hep-th].
\bibitem{FPW}
  S.~Folkerts, A.~Pritzel and N.~Wintergerst,
  arXiv:1107.3157 [hep-th].
\bibitem{Kiritsis:2006hy}
  E.~Kiritsis,
  JHEP {\bf 0611} (2006) 049
  [hep-th/0608088].
\bibitem{Aharony:2006hz}
  O.~Aharony, A.~B.~Clark and A.~Karch,
  Phys.\ Rev.\ D {\bf 74} (2006) 086006
  [hep-th/0608089].
\bibitem{vDV}   
  H.~van Dam and M.~J.~G.~Veltman,   
  Nucl.\ Phys.\  B {\bf 22} (1970) 397.   
\bibitem{HHV}
  S.~F.~Hassan, S.~Hofmann and M.~von Strauss,
  JCAP {\bf 1101} (2011) 020
  [arXiv:1007.1263 [hep-th]].
\bibitem{Nunes}
  F.~C.~P.~Nunes and G.~O.~Pires,
  Phys.\ Lett.\  B {\bf 301} (1993) 339.
\end{thebibliography}
\end{document}